\documentclass[conference]{IEEEtran}

\IEEEoverridecommandlockouts
\usepackage{amsmath,amssymb,amsfonts}
\usepackage{cite}
\usepackage{algorithmic}
\usepackage{graphicx}
\usepackage{textcomp}
\usepackage{xcolor}
\usepackage{svg}
\usepackage{footmisc}
\usepackage{multirow}
\usepackage{pifont}   % pour \ding{54}
\usepackage{subcaption}
\usepackage{diagbox}
\usepackage{caption}

\def\BibTeX{{\rm B\kern-.05em{\sc i\kern-.025em b}\kern-.08em
    T\kern-.1667em\lower.7ex\hbox{E}\kern-.125emX}}

\begin{document}

\title{Explainable Anomaly Detection in 5G Networks through Continuous Time Graphs}

\author{
  \IEEEauthorblockN{Thomas Hoger}
  \IEEEauthorblockA{
    \textit{LAAS - CNRS, Université de Toulouse} \\
    Toulouse, France \\
    thomas.hoger@laas.fr
  }
  \and
  \IEEEauthorblockN{Philippe Owezarski}
  \IEEEauthorblockA{
    \textit{LAAS - CNRS, Université de Toulouse} \\
    Toulouse, France \\
    owe@laas.fr
  }
}

% \and
% \and
% \and
% \and

\maketitle

\begin{abstract}
5G networks increasingly resemble distributed systems in their design. It then becomes crucial to move beyond traditional network-based detection methods and to explore alternative approaches. In this work, we propose a formalization of 5G network exchanges as a dynamic graph, on which we apply techniques from the field of graph-based learning to detect anomalies effectively. This representation allows for fine-grained analysis of packet-level semantic content, while simultaneously capturing sequential and topological contexts, both of which are essential for identifying sophisticated and multi-stage attacks. To assess the effectiveness of this formalism, we evaluate it with the most popular dynamic graph models, demonstrating its potential for 5G anomaly detection.
\end{abstract}


\section{Introduction}

% La 5G change les réseaux mobiles
Beyond the improvements in service quality enabled by physical innovations, 5G  also introduces a major paradigm shift for mobile network architectures. While the core network was traditionally deployed on specialized hardware and relied on dedicated protocol stacks, it is now virtualized and software-driven, with independent network functions interacting through standardized interfaces. This paradigm shift brings 5G architectures closer to distributed systems, while making the network more flexible and dynamic.

% Sémantique -> packet level -> IDS plus suffisant pour les attaques high level
Yet, these benefits also come at the cost of an increased attacking surface. Indeed, 5G remains vulnerable to attacks affecting traditional networks but also introduces novel attack vectors specific to its new ecosystem\cite{pfcp,pfcp_in_gtp,spoof_gtp, 5gad}. As these attacks rely on application-layer payloads, traditional IDSs based on quantitative flow analysis are no longer sufficient. This highlights the need for techniques that consider network traffic at the packet level.

% Dimensions
We identify three dimensions for characterizing these attacks. The semantic dimension captures the intrinsic content of packets. However, attacks can also manifest through anomalies in the surrounding context. This context can be sequential, when messages from the same entity are repeatedly exchanged, or topological, when the communication pattern between entities is abnormal, as in a Man-in-the-Middle (MITM) attack.

% Manque dans SOTA
These dimensions are complementary and should be considered jointly to detect novel 5G threats. Yet, their simultaneous consideration remains largely absent from the state of the art. For instance, AutoGuard, ADSeq, and 5GCGuard focus exclusively on the sequential dimension, which they analyze using RNNs \cite{autoguard,adseq,5GCGuard}. Similarly, Kale et al. \cite{ensemble} consider only the semantic dimension by analyzing packet content with CNNs. Other approaches, as Lam et al. \cite{cnn}, combine two dimensions by feeding the output of a CNN into an RNN, capturing both the semantic and sequential dimensions while overlooking the topological one.

% GSAD
To the best of our knowledge, GSAD \cite{gsad} is the only approach that jointly considers the three dimensions. It combines a CNN for semantics, an RNN for sequences, and a graph neural network (GNN) for interactions between entities. However, its semantic analysis remains limited by the CNN, which simply aggregates byte values without explicitly interpreting their meaning, potentially missing subtle value shifts. Furthermore, the sequential use of three separate models limits their ability to capture correlations across dimensions. These limitations motivate the use of a unified representation that can jointly capture these dimensions rather than processing them through separate models.

% Temporal graph
These limitations call for a more integrated representation of network interactions. Dynamic graphs provide such a framework, naturally representing semantic, sequential, and topological information within the same structure. We therefore represent 5G communications as Continuous-Time Dynamic Graphs (CTDGs) and experimentally evaluate this approach using state-of-the-art machine learning models within a recent framework for benchmarking dynamic graph learning methods. Throughout this paper, we use the terms \textit{formalism} to refer to the proposed graph representation and \textit{model} exclusively to refer to the machine learning models applied to it.

% Interpretabilité important
Beyond capturing the topological dimension, graph formalizations also offer a visual and interpretable view of network interactions. The proposed formalization therefore also provides a high degree of explainability, which is particularly valuable in operational settings, where anomalies may result from novel but legitimate behavior.

% Overview
The paper is organized as follows: Section \ref{objectives} introduces the key contextual elements that define the scope of our problem. Section \ref{related_works} reviews related work on graph neural networks. Section \ref{formalization} presents our formalization, while Section \ref{methodology} describes our anomaly detection methodology. Section \ref{experiments} details the models and datasets used to validate our approach. Finally, Section \ref{conclusion} concludes the paper by summarizing our contributions and outlining future work.

\section{Preliminiaries}
\label{objectives}

\begin{figure}[!t]
\centering
\includegraphics[width=\linewidth]{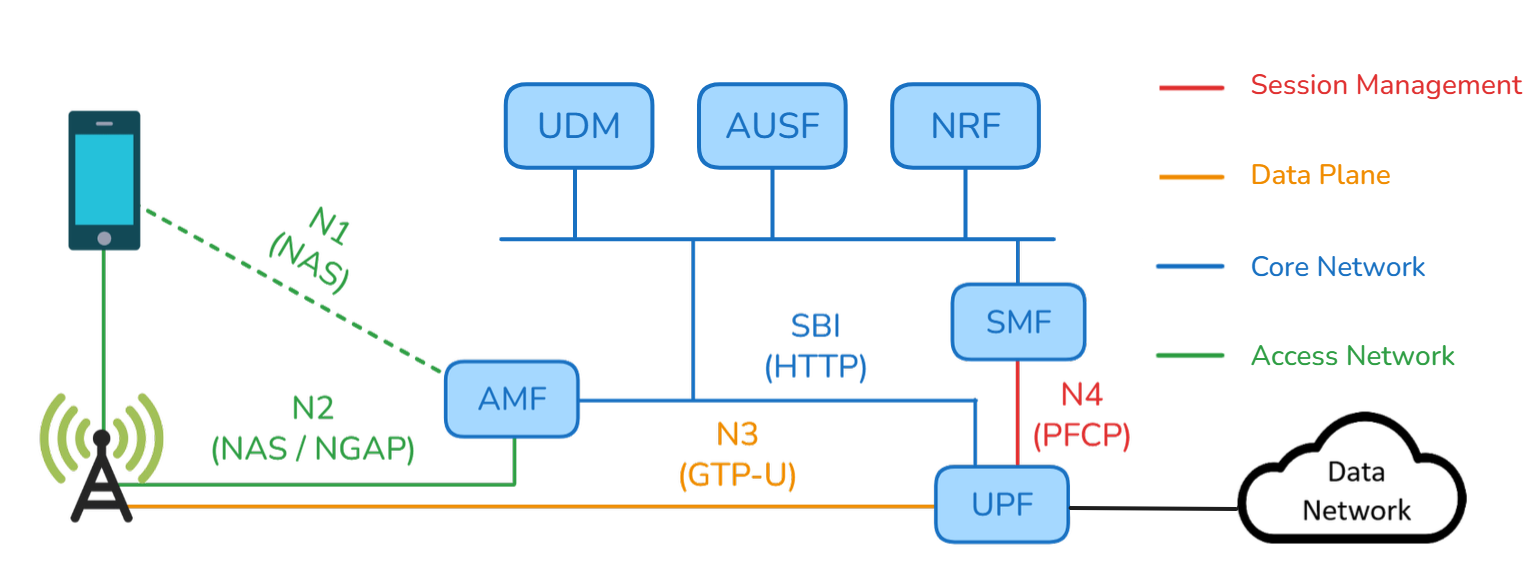}
\caption{Non-roaming 5G architecture and key protocols}
\label{fig:architecture}
\end{figure}

\subsection{Scope of the study}
A 5G architecture consists of User Equipment (UE) connected to base stations (gNBs), which form the radio access network (RAN). This access network interacts with the 5G Core Network (CN), which is composed of independent network functions (NFs), each fulfilling a specific role and communicating through HTTP API calls. Among other functions, the CN uses the Packet Forwarding Control Protocol (PFCP) to establish data-plane tunnels between the gNBs and the Data Network (DN), which carry user traffic using the GTP-U protocol.

The 5G CN is designed to be highly modular, allowing network operators to deploy different types of NFs depending on their requirements. It is therefore impractical to consider all possible network configurations. In this work, we focus on the set of NFs and application-layer protocols required for the basic operation of a 5G network, as illustrated in Figure~\ref{fig:architecture}. This scope provides a manageable setting for our proof of concept while remaining extensible. Indeed, additional NFs and protocols can be incorporated into our approach without requiring changes to the proposed method.

\subsection{Threat model}
The vast majority of state-of-the-art attacks require an entry point into the 5G network to be established beforehand, either by compromising an existing machine or by integrating a new hostile actor. However, there are many ways to achieve this. For example, an attacker can exploit vulnerabilities in slices\cite{slice2} to pivot to machines on a network to which they previously had no access. Since network functions are virtualized, it is also possible to reach them via their host\cite{docker}. Finally, an attacker could take advantage of the proximity between gNB and users to physically access one of them and take control of it. We therefore make the preliminary assumption that an attacker has managed to break into the network by taking control of one or more UE, gNB, or CN functions. 

\subsection{User Plane vs. Control Plane Traffic}
5G networks carry two types of traffic: user traffic and control traffic. User traffic consists of data exchanged between users and the Internet, and typically accounts for most of the network traffic volume. It represents the user actual use of the connectivity and is encrypted with TLS. From the network perspective, it consists simply of messages forwarded by the 5G core and does not exhibit any characteristics specific to 5G.

On the other hand, control traffic consists of messages used to establish communication tunnels and operate the 5G network. Unlike user traffic, it relies on behaviors and protocols specific to 5G and mobile networks. These protocols and behaviors introduce specific attack vectors that are therefore of particular interest from a security perspective. Although encrypted, control traffic can be decrypted and inspected by the operator, who controls the network equipment and holds the necessary cryptographic secrets. This makes control traffic particularly relevant for security analysis.

However, several works in the state of the art do not distinguish between these two types of traffic, considering either attacks \cite{5g_nidd} or benign traffic \cite{5gad} from the user plane, despite its lack of characteristics specific to 5G. Thus, in this work, we focus exclusively on benign and anomalous traffic within the 5G control plane.

%\subsection{Security Assumptions}
%In this study, we adopt the perspective of the operator managing the 5G network and consider that the control traffic it intercepts is readable in plain text. Indeed, even if communication between NFVs is encrypted point-to-point by the TLS protocol, the operator has control over each machine in the CN and can therefore decrypt messages at will. Since the attacker does not control the network, we consider that they are unable to decrypt the traffic retrieved by passive listening. We therefore limit our study to attacks requiring the active participation of an attacker. 
\section{Related works}
\label{related_works}

% Intro
Graphs are inherently well suited to represent relationships, which is why they have been used in works such as GSAD \cite{gsad} to capture the topological dimension. In this section, we review traditional GNNs as well as their temporal extensions, which enable them to represent the evolution of these relationships over time.

% Fonctionnement général + origine
GNNs perform neighborhood-based aggregation, enriching each node representation with information from its local context. Most modern GNNs follow the message passing paradigm introduced by Gilmer et al. \cite{mpnn}, which consists of three steps: constructing messages from neighboring node representations, aggregating them, and updating the node representations.

% Static -> Dynamic Discret
Several extensions have been proposed, including Graph Attention Networks (GATs) \cite{gat}, which use attention to weight neighboring information, and GraphSAGE \cite{graphsage}, which supports inductive learning on unseen nodes. However, these architectures are primarily designed for static graphs, whereas our data evolves continuously over time. We therefore consider dynamic graph models, which explicitly incorporate temporal information. These models can be broadly divided according to their temporal granularity. Discrete-Time Graph Networks (DTGNs), such as EvolveGCN \cite{evolvegcn} and GSAD \cite{gsad}, represent the graph as a sequence of discrete snapshots and combine standard GNN operations with recurrent models to capture temporal evolution.

% Dynamic Continu
The second category consists of Continuous-Time Graph Networks (CTGNs), which operate at a finer temporal granularity than discrete-time approaches. Rather than representing the graph through fixed snapshots, they represent the graph as a sequence of events, such as node or edge additions, deletions, and updates. CTGNs can be further divided into two families. The first family consists of temporal neighborhood methods, such as TGN~\cite{tgn}, DyRep~\cite{dyrep}, and JODIE~\cite{jodie}, which maintain a latent memory for each observed node, and update it as new events occur. The second family consists of temporal embedding methods, such as GraphMixer~\cite{graphmixer}, CAWN~\cite{cawn}, TCL~\cite{tcl}, and DyGFormer~\cite{dygformer}. In contrast to the first family, the second one do not explicitly store node states in the memory and instead derive temporal representations directly from the observed interactions. Memory-based approaches retain a condensed history of past interactions, providing richer contextual information, while memory-free approaches require less memory and computational resources.

% Conclusion
Each of these CTGNs differs in how it operates and in the advantages it offers, as demonstrated by the comparative benchmark of Feng et al.~\cite{survey_gnn}. In this work, we evaluate several representative and widely used models on our problem to determine which is best suited to 5G anomaly detection.

\section{Proposed Graph Formalization}
\label{formalization}
The main contribution of this work lies in the introduction of two complementary graph representations of the same underlying network activity. The static representation provides a highly interpretable view of the graph structure, while the dynamic representation enables the use of CTGNs to jointly capture the three dimensions during learning.

\begin{figure}[ht]
\centering
\begin{subfigure}[b]{0.35\linewidth}
    \centering
    \includegraphics[width=\linewidth]{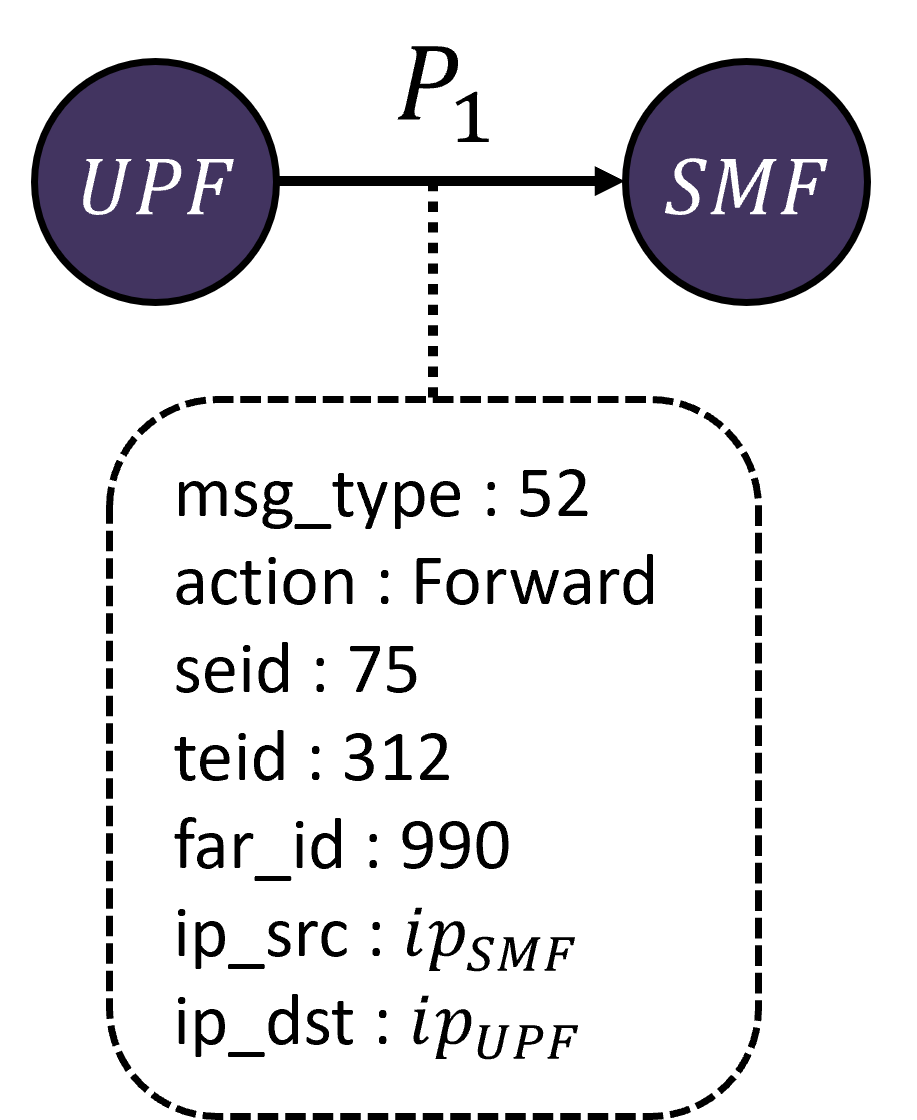}
    \caption{GSAD \cite{gsad}}
    \label{fig:form_gsad}
\end{subfigure}
\hfill
\begin{subfigure}[b]{0.6\linewidth}
    \centering
    \includegraphics[width=\linewidth]{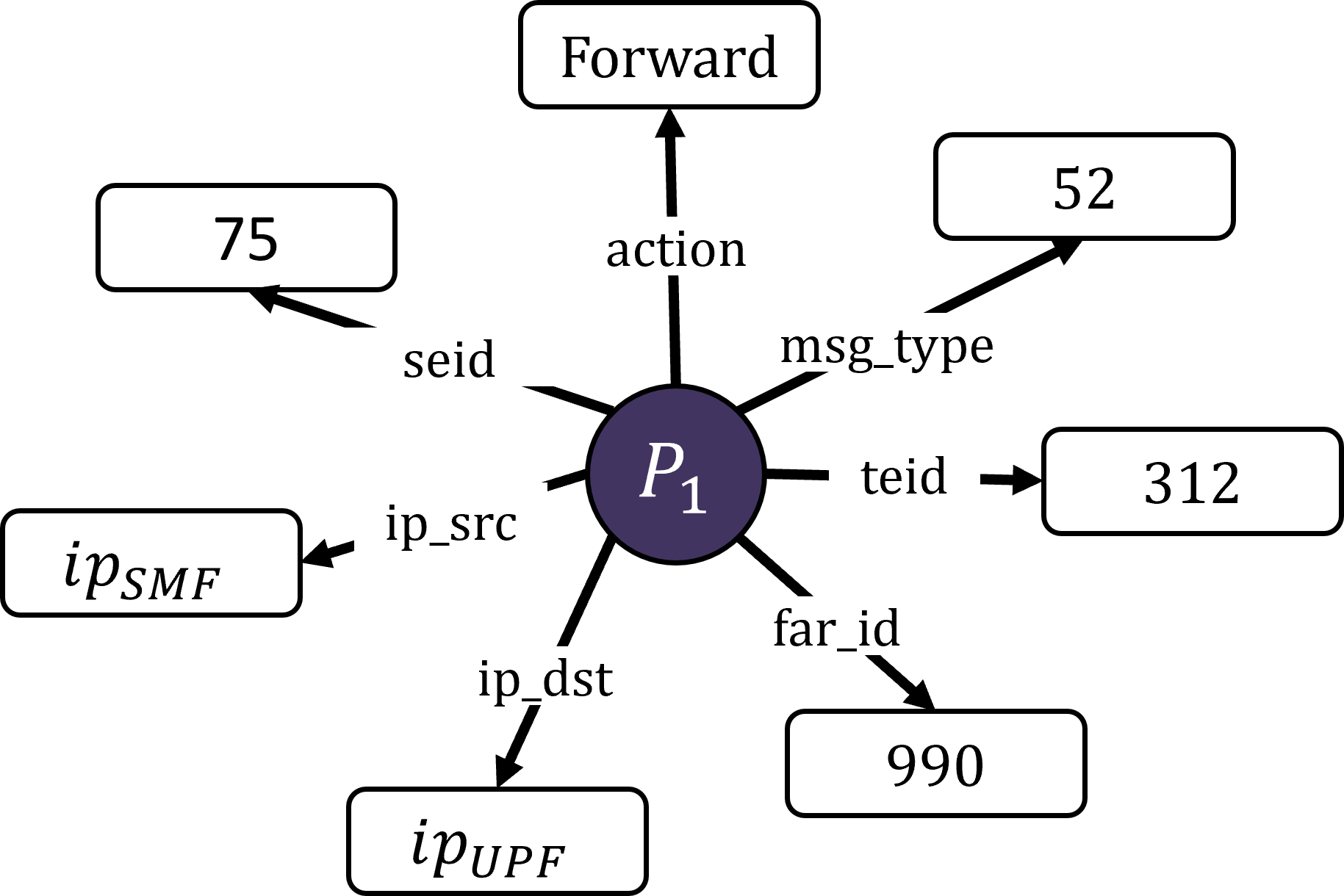}
    \caption{Proposed formalism}
    \label{fig:form_us}
\end{subfigure}
\caption{Comparison between the GSAD formalism and our proposed formalism for a packet $P_1$ exchanged between an UPF and an SMF}
\label{fig:difference}
\end{figure}

\begin{figure}[ht]
\centering

\begin{subfigure}[b]{0.7\linewidth}
    \centering
    \includegraphics[width=\linewidth]{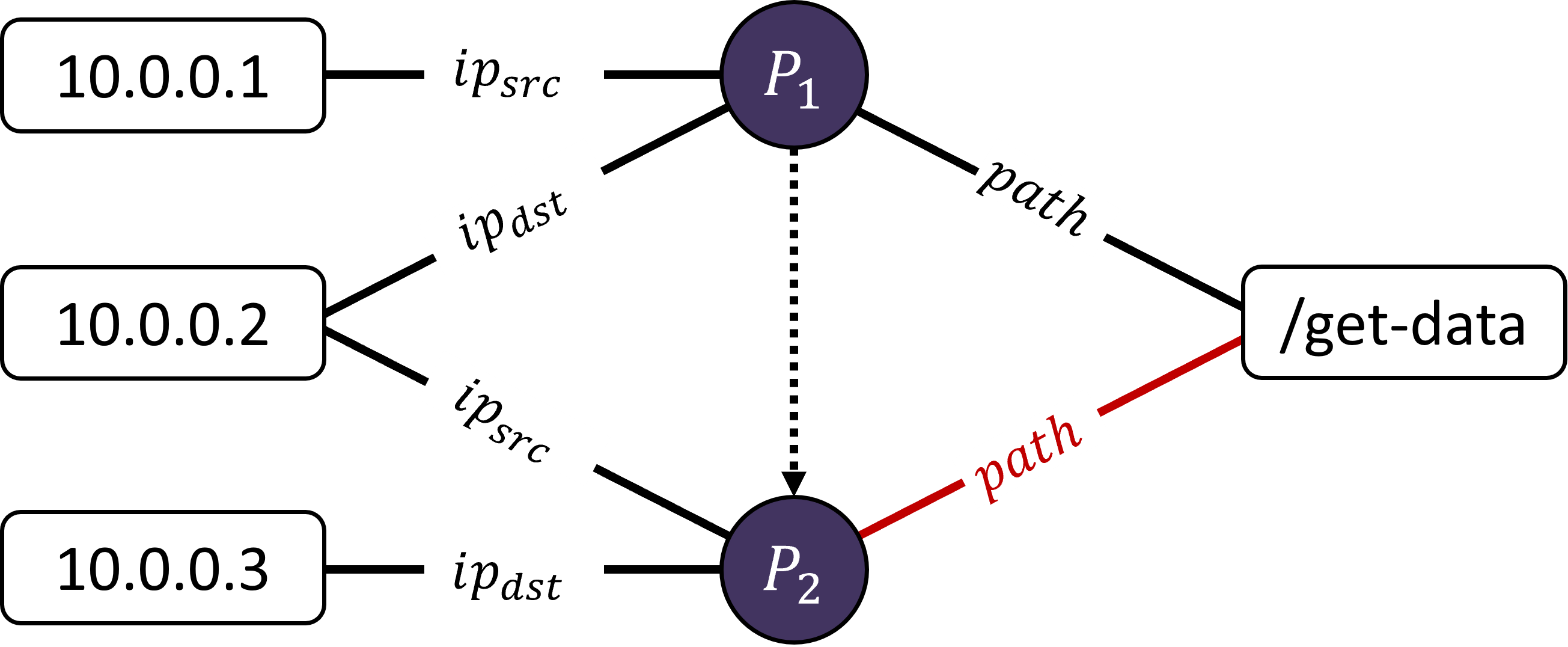}
    \caption{Topological anomaly (MITM)}
    \label{fig:anomaly_topological}
\end{subfigure}

\vspace{0.5em}

\begin{subfigure}[b]{0.7\linewidth}
    \centering
    \includegraphics[width=\linewidth]{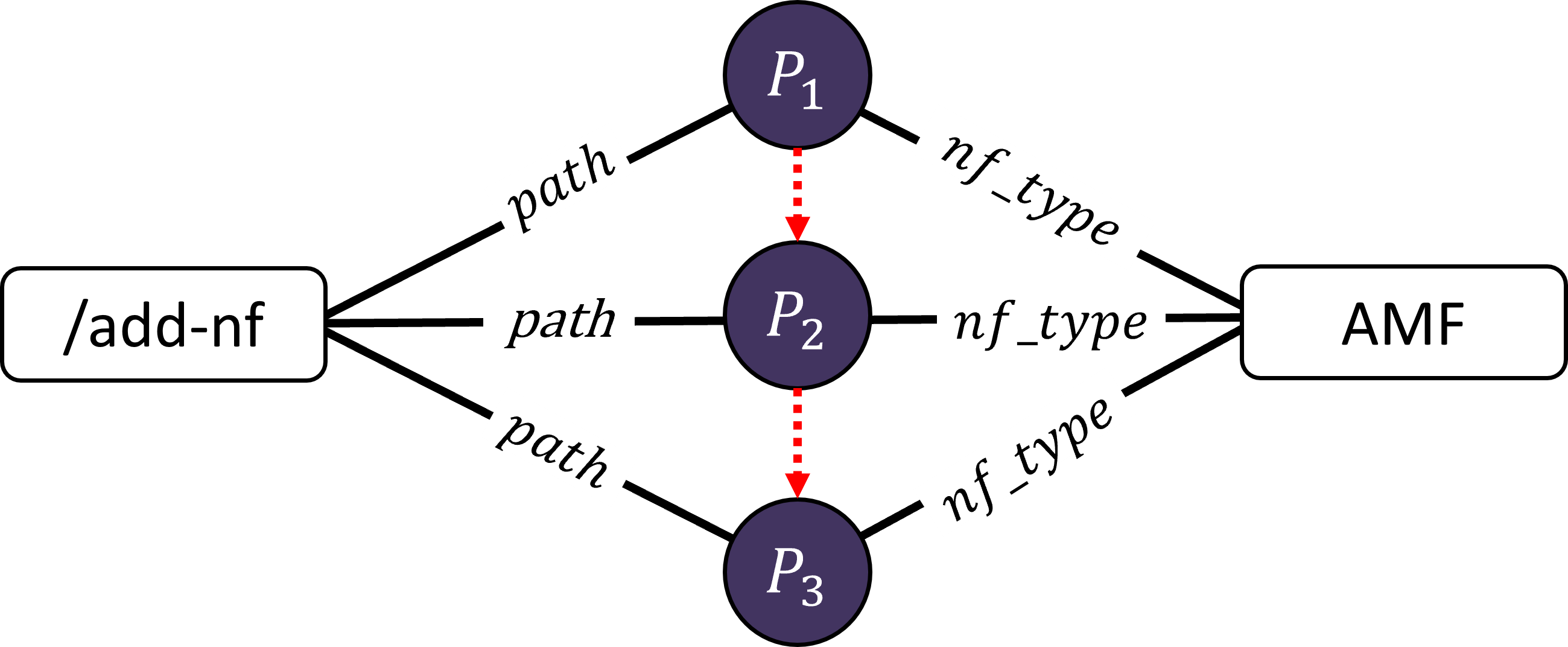}
    \caption{Sequential anomaly (Flooding)}
    \label{fig:anomaly_sequential}
\end{subfigure}

\vspace{0.5em}

\begin{subfigure}[b]{0.7\linewidth}
    \centering
    \includegraphics[width=\linewidth]{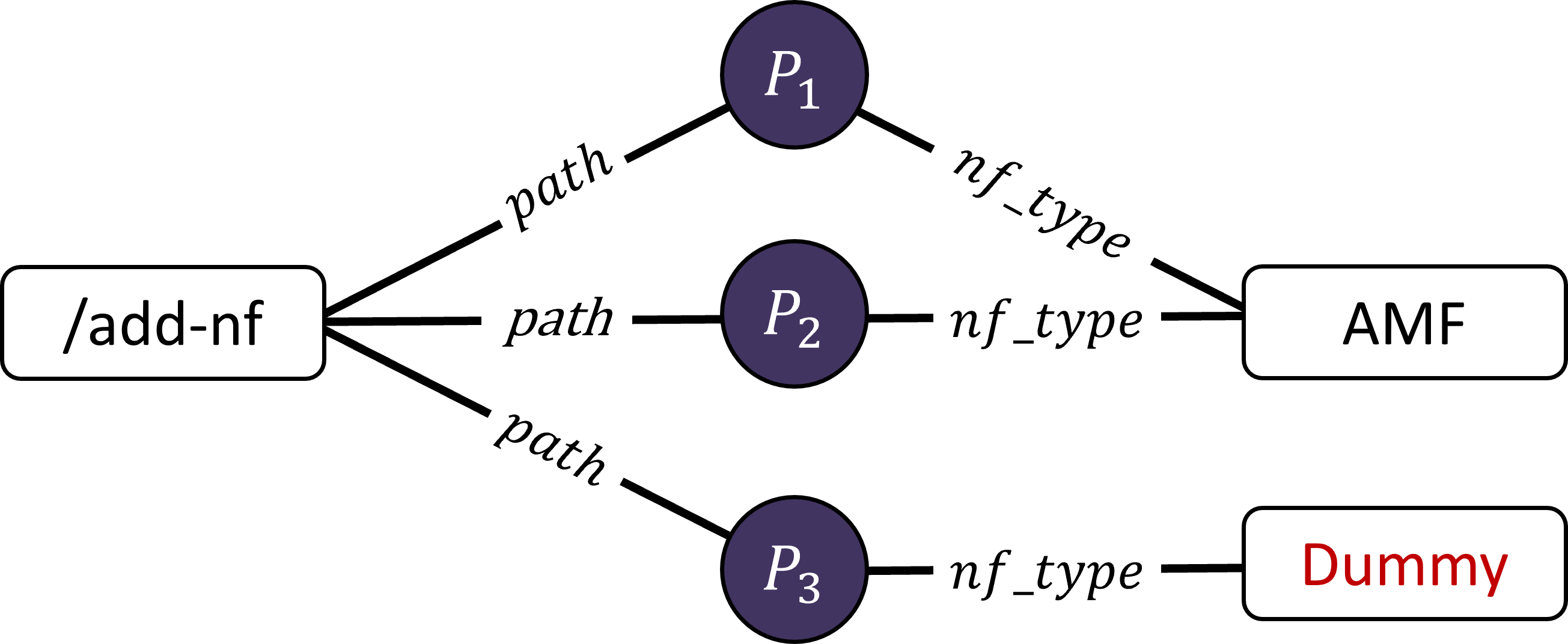}
    \caption{Semantic anomaly (Fuzzing)}
    \label{fig:anomaly_semantic}
\end{subfigure}

\caption{Anomalies from the three dimensions represented as static graphs. Indicators of anomalous interactions are highlighted in red.}
%Examples of anomalies represented as graphs. (a) A topological anomaly corresponding to an abnormal API call redirection, represented by the red edge. (b) A semantic anomaly corresponding to an abnormal parameter value, represented by the red node. (c) A sequential anomaly corresponding to an abnormal sequence of identical benign packets, with the anomaly represented by the red sequence dependency.}
\label{fig:anomalies}
\end{figure}

\subsection{Static Formalization}

% Ce que fait GSAD
As illustrated in Figure~\ref{fig:difference} (a), GSAD \cite{gsad} represents a packet and its payload as a single edge $P_1$ connecting two different network entities, namely the SMF and the UPF here. In this approach, every feature contained in the packet, such as IP addresses or URL paths are aggregated into the edge indiscriminately, which requires the use of methods such as CNNs to extract and process these features.

% Ce qu'on fait nous
In contrast, as proposed in Figure~\ref{fig:difference} (b), we decompose each packet into its constituent features. Each packet $P_x$ is represented by a \textit{central node}, while each of its features is represented by an individual \textit{feature node}. The \textit{central node} is connected to each \textit{feature node} through an edge labeled with the corresponding feature name, forming a subgraph that explicitly represents the packet and its content.

% Création du graph
Importantly, feature nodes are shared across packet subgraphs. When a feature value already exists in the graph, we reuse the corresponding node rather than creating a new one. This mechanism naturally connects packets that share common features. Consequently, processing a packet always adds one new \textit{central node}, while new \textit{feature nodes} are created only for feature values that have not previously appeared in the graph. The number of newly added \textit{feature nodes} corresponds to the number of features in the packet that are not already present in the graph.

% Fleche pointillé
To capture temporal dependencies, we augment the graph with a directed connection between successive packets, represented as a dashed edge. This temporal connection explicitly represents the order in which packets occur and makes the temporal dependencies between them visible in the graph. This results in representations, such as the one shown in Figure \ref{fig:anomalies}, which illustrates examples of the three dimensions of anomalies. In each case, the anomalous dimension can be readily identified from the information highlighted in red. 

% Explication figure
Figure (a) illustrates a topological attack involving a MITM redirection, where a legitimate packet $P_1$ is copied into a new packet $P_2$ and forwarded to another destination. Figure (b) illustrates a sequential anomaly, where the packets themselves are legitimate but their order of arrival is abnormal. Figure (c) illustrates a semantic anomaly in packet $P_3$, whose \textit{nf-type} field contains an invalid value. By comparing $P_3$ with similar requests, the model can detect that its content deviates from the expected pattern, without requiring explicit knowledge of the \textit{nf-type} field.

% Conclusion
Because of the packet decomposition, the proposed formalism enables attributes from different packets to be directly connected through shared characteristics, rather than relying solely on the identities of communicating entities, as in previous graph-based representations. This allows the model to capture richer relationships between packet features and leverage the GNN to capture the semantic dimension directly, without requiring a separate CNN to process these features. Moreover, by explicitly representing each feature in the graph, our approach makes it possible to identify the specific parameters contributing to anomalous behavior, providing an inherently interpretable representation.

\begin{figure}[ht]
\centering
\includegraphics[width=\linewidth]{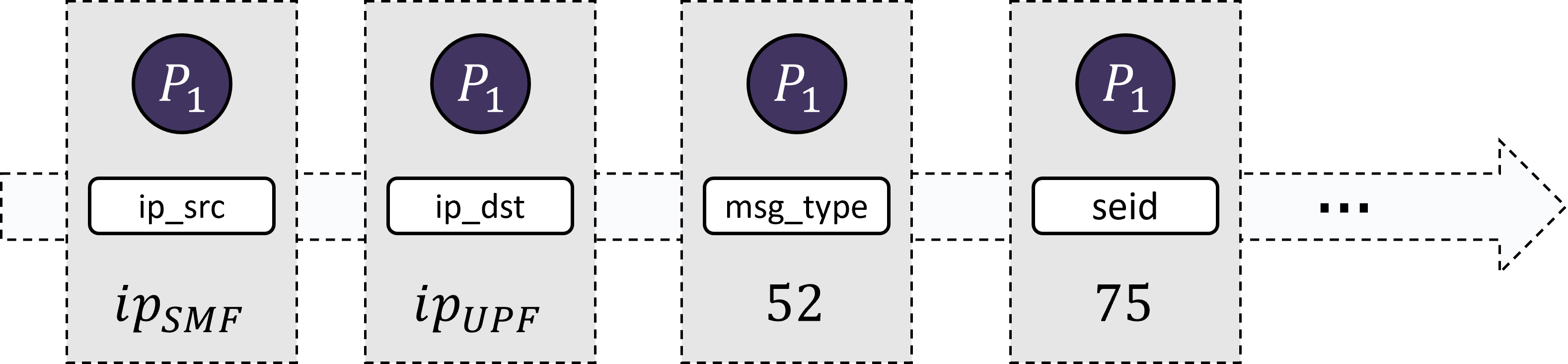}
\caption{Packet represented as a sequence of events}
\label{fig:seq}
\end{figure}

\begin{figure*}[h]
\centering

% --- Première ligne ---
\begin{subfigure}[b]{0.2\linewidth}
    \centering
    \includegraphics[width=\linewidth]{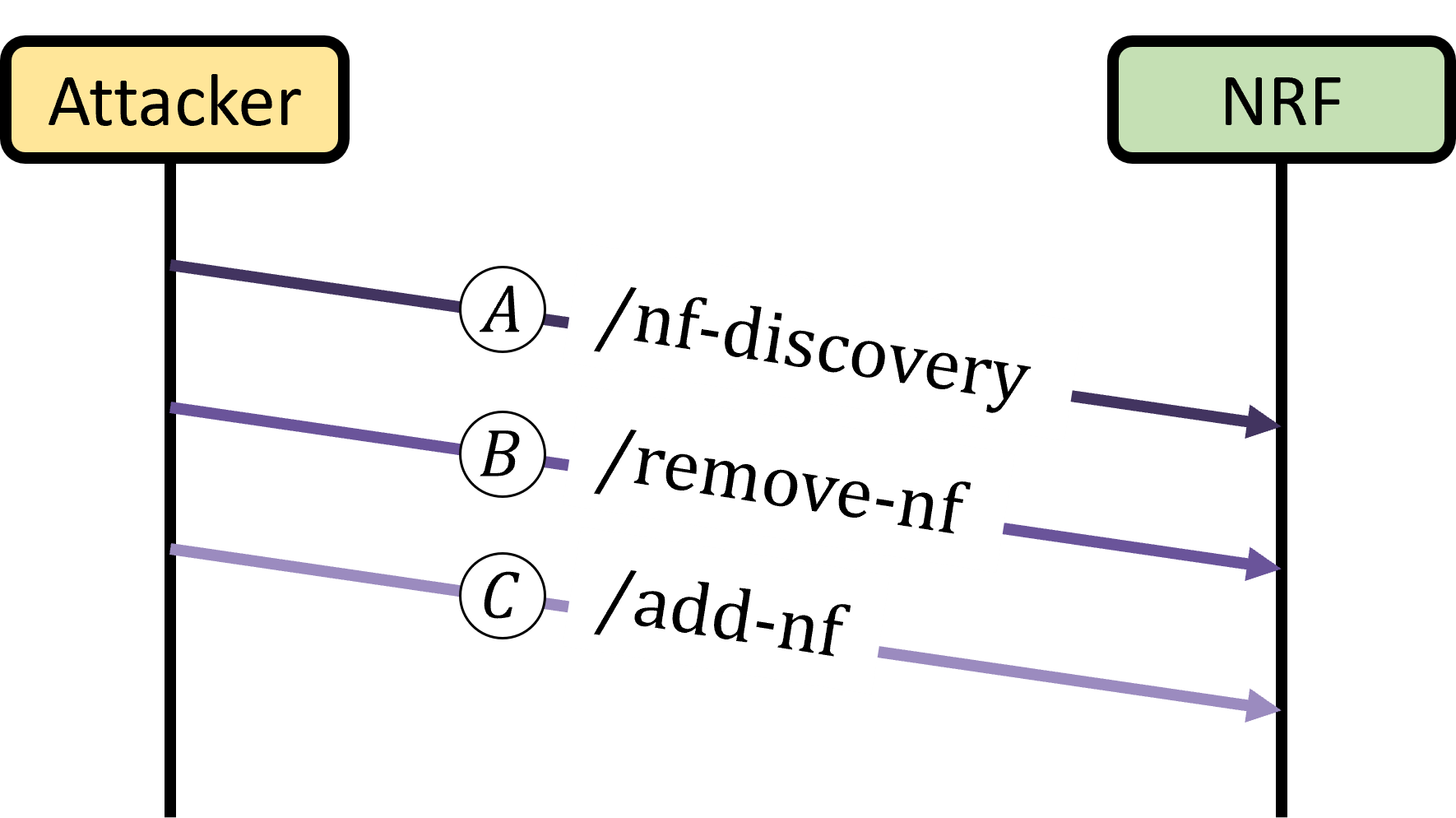}
    \caption{Sequence of 3 packets}
\end{subfigure}
\hfill
\begin{subfigure}[b]{0.25\linewidth}
    \centering
    \includegraphics[width=\linewidth]{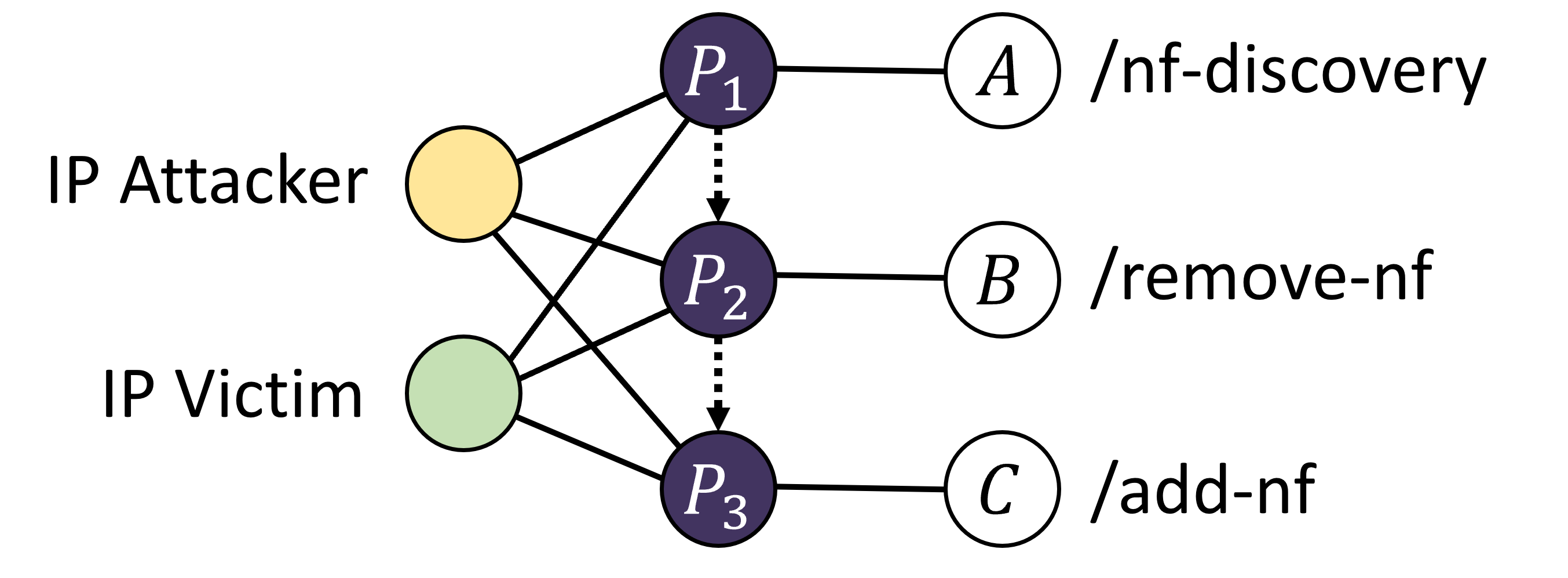}
    \caption{Static Graph representation}
\end{subfigure}
\hfill
\begin{subfigure}[b]{0.45\linewidth}
    \centering
    \includegraphics[width=\linewidth]{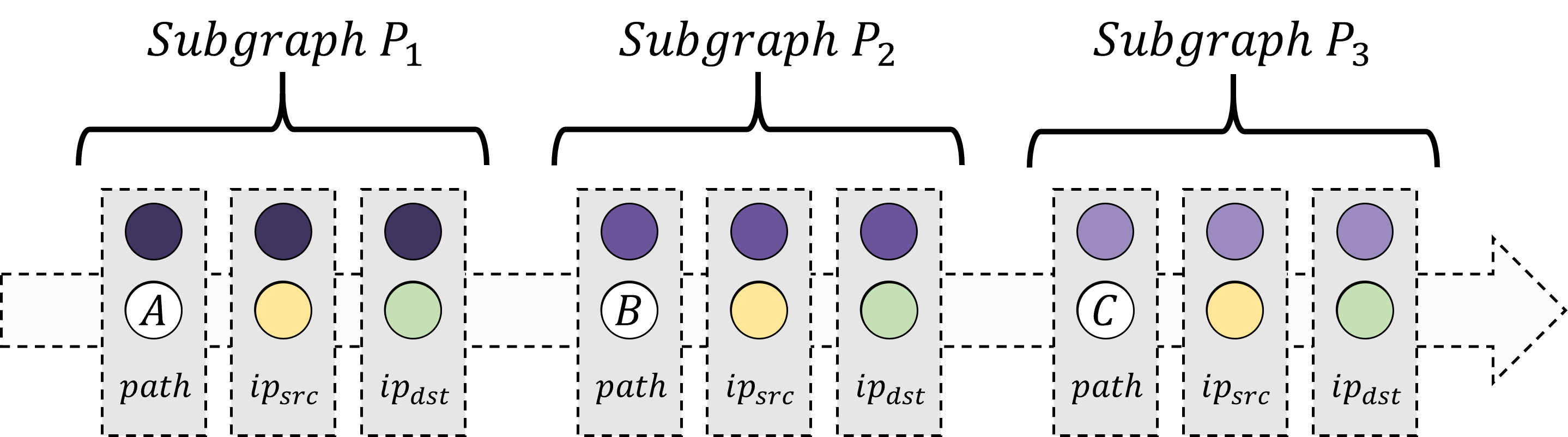}
    \caption{Continuous Time Graph representation}
\end{subfigure}

\caption{Different representations of the same sequence of exchanges between an attacker and the NRF}
\label{fig:graph_continu}
\end{figure*}

\subsection{Dynamic Formalization}

% Nécessite de dynamique
The static formalism presented above is well suited for providing a visual representation of the data, making it particularly useful for human interpretation after inference. However, its use for learning would require a static GNN, which is less suited to capturing temporal dependencies. We therefore adopt a Continuous Time Graph (CTG) formalism that will be used for both learning and inference. In this formalism, each feature is represented as an event rather than a static edge. This event is defined by a central node, a feature node, and the corresponding feature name.

% Explication plus détaillé
Each packet therefore forms a set of events corresponding to its subgraph in the static representation, as illustrated in Figure \ref{fig:seq}, which presents the same packet content shown in Figure \ref{fig:difference}, this time represented as a sequence of events. 

Since all features of a given packet are observed simultaneously, their ordering within the set is not meaningful, as long as it remains consistent across all packets. Meaningful temporal patterns instead arise from the temporal ordering of sets generated by different packets in the sequence.

% Figure
Figure \ref{fig:graph_continu} summarizes and illustrates the two formalisms proposed in this work using a sequence of three packets sent by an attacker to an NRF. The three packets are first represented in their original order in Figure (a). In Figure (b), they are formalized into a static graph, where each packet generates its own central node and corresponding features, forming a subgraph interconnected with one another. Figure (c) shows the corresponding CTG, where the three packets form three sets of events, each containing one event per feature. Events within each set are ordered consistently across packets.

% Conclusion
This dynamic representation thus preserves the semantic and topological information captured by the static graph while explicitly encoding the temporal evolution of network activity. It provides a unified representation of the three anomaly dimensions and forms the basis for applying temporal graph learning methods to their detection.

\section{Anomaly Detection Methodology}
\label{methodology}

\begin{figure}[ht]
\centering
\includegraphics[width=\linewidth]{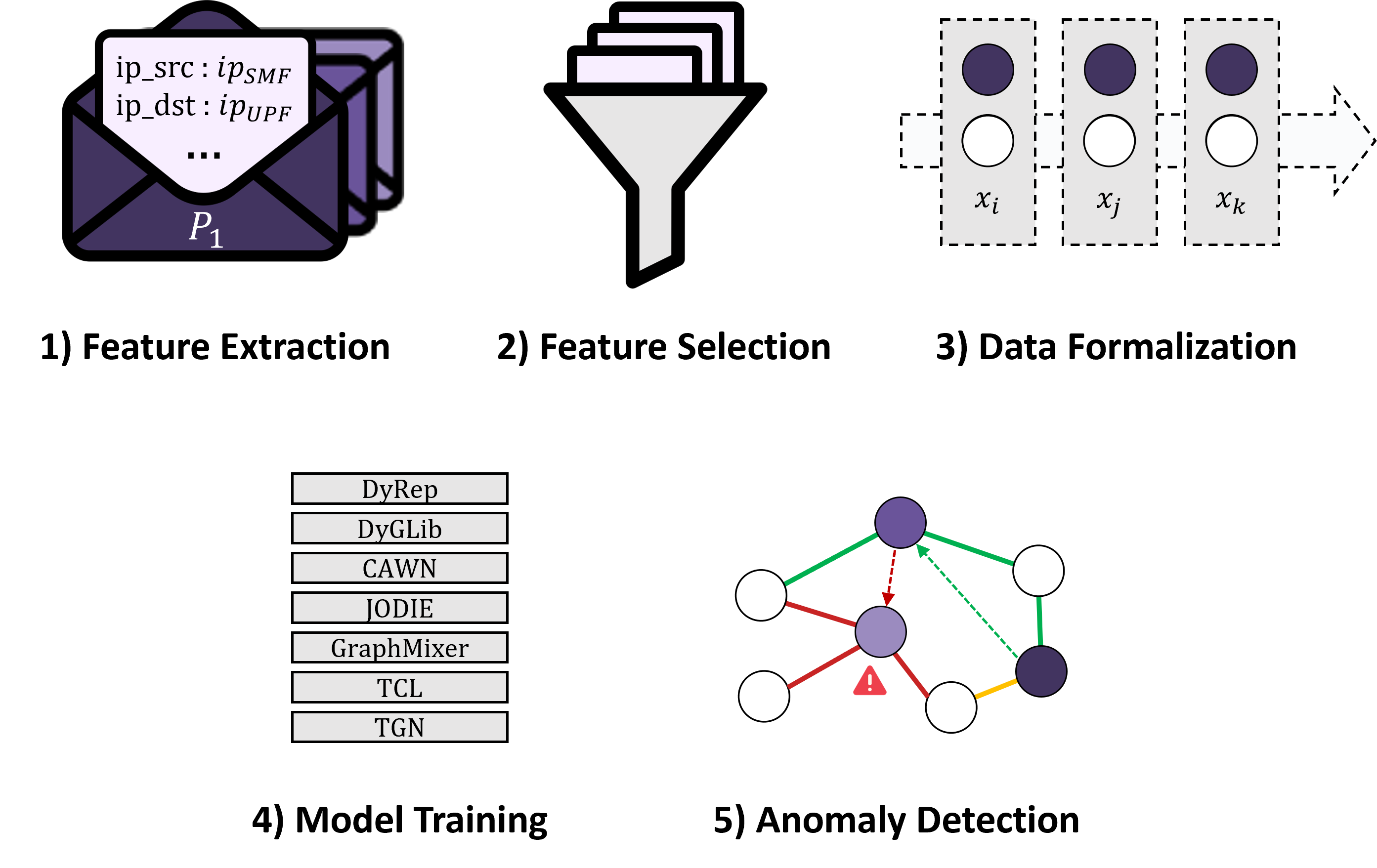}
\caption{Overview of the proposed anomaly detection pipeline}
\label{fig:pipeline}
\end{figure}

This section presents the different stages of our pipeline, from data processing and feature extraction to model training and inference. Figure \ref{fig:pipeline} illustrates the overall workflow of the proposed approach.

\textbf{Feature extraction} : We process network packets from PCAP traces sequentially and use the Scapy library to extract the payloads of the application-layer protocols. For each packet, we store each of the extracted features in a structured JSON representation.

\textbf{Feature selection}: To avoid introducing noise and hindering the model ability to learn, we empirically selected 41 features that we identified as characteristic of the attacks considered in our dataset.

%considered as the most relevant for identifying the attacks present in our dataset. The selected features, together with the attacks they can help detect, are reported in Table~\ref{tab:feature_attack_mapping}.

% \textbf{Feature Encoding} : Features are encoded using one-hot encoding to obtain numerical vectors suitable for the machine learning model. These vectors are subsequently used to represent the semantic information associated with the graph edges.

\textbf{Data formalization} : The resulting JSON packets are then transformed into a sequence of events following the formalization presented in Section~\ref{formalization}. Each event contains a central and feature node ID, as well as the ID of the edge connecting them. For each of those 3 IDs, we maintain a corresponding vector $x_i$. Edge vectors are initialized from the one-hot encoding of the associated feature name, while node vectors are initialized to zeros and progressively acquire meaningful representations through successive message-passing operations. In addition to these vectors, we maintain metadata for each event, including the packet ID and its ground-truth label.

\textbf{Model Training} : We train state-of-the-art GNNs for supervised link prediction using the models implemented in the DyGLib library\footnote{https://github.com/yule-BUAA/DyGLib}, which provides a standardized framework. The evaluated models are those presented in the related work, namely TGN\cite{tgn}, GraphMixer\cite{graphmixer}, DyGFormer\cite{dygformer}, CAWN\cite{cawn}, TCL\cite{tcl}, JODIE\cite{jodie}, and DyRep\cite{dyrep}, providing a representative selection of popular CTGNs. Using DyGLib as an unified framework ensures consistency across models and avoids implementation-related variability that could arise from using different sources. %In addition to detection performance, we measure the execution time of each model to assess the practical feasibility of the different approaches in an operational setting.

\textbf{Anomaly detection} : Once the model has been trained, it produces probability scores indicating the likelihood of an edge existing between a central node and a feature node, based on how closely the observed interaction resembles patterns previously encountered during training. This enables anomalies to be detected along one or multiple detection dimensions. The resulting probability scores can then be used to represent the detected anomalies as a static graph, as presented in Figure~\ref{fig:anomalies}.

\section{Experiments}
\label{experiments}

\subsection{Dataset}

% Seuls dataset sont sur userplane
To the best of our knowledge, only three papers from the state of the art present datasets specifically designed as 5G datasets for anomaly detection \cite{5g_nidd,pfcpdataset,5gad}. However, their benign traffic consists exclusively of user-plane traffic, which cannot provide an appropriate baseline for comparison with attacks targeting the 5G control plane.

% Autres papiers detect font leur propre dataset
In the absence of a satisfactory public dataset, anomaly detection studies generally have to generate their own training data \cite{gsad,prov5g,adseq,5GCGuard}. However, the data generation process is rarely described in sufficient detail, and none of these works provides an analysis of dataset quality to identify potential biases or verify that the proposed attacks are comparable in nature to benign traffic. Among these works, 5GCGuard \cite{5GCGuard} is, to the best of our knowledge, the only one providing tools to reproduce the data generation process. However, only a preprocessed version of the generated dataset is publicly available, without access to the original raw captures, which limits reproducibility. 

% On fait le notre
Given the lack of a publicly available dataset that has reached a consensus within the community, we therefore constructed our own dataset using UERANSIM and free5GC, as in several of the aforementioned works. Particular attention was paid to avoiding potential biases identified in \cite{badsmell}. We explicitly defined our scope and the protocols considered in our study, namely HTTP/2, GTP-U, PFCP, and IP, to avoid highly dependent features. Although IP addresses are used to identify communicating entities, attacker identities are randomized to prevent the model from exploiting them as shortcuts for attack detection. We also avoid temporal biases by intertwining benign and attack traffic throughout the capture, with both generated in randomized order.This resulted in a four-hour traffic capture containing approximately 1,200,000 benign messages and 130,000 attack messages, which were subsequently labeled.

\begin{table}[t]
\centering
\begin{tabular}{|l|c|c|c|}
\hline
\textbf{Attack name} & \textbf{Semantic} & \textbf{Sequential} & \textbf{Topological} \\ \hline

\multicolumn{4}{|c|}{\textbf{SBI (HTTP/2)}} \\ \hline
Applicative MiTM & & & \ding{54} \\ \hline
Applicative Scan \cite{5gad} & & \ding{54} & \\ \hline
API Fuzzing & \ding{54} & \ding{54} & \\ \hline

\multicolumn{4}{|c|}{\textbf{N4 (PFCP)}} \\ \hline
Establishment Flood \cite{pfcp} & & \ding{54} & \\ \hline
Deletion Flood \cite{pfcp} & & \ding{54} & \\ \hline
SEID Fuzzing & \ding{54} & \ding{54} & \\ \hline
Modify Drop \cite{pfcp} & \ding{54} & \ding{54} & \\ \hline
Modify Duplicate \cite{pfcp} & \ding{54} & \ding{54} & \\ \hline

\multicolumn{4}{|c|}{\textbf{N3 (GTP-U)}} \\ \hline
Uplink Spoofing \cite{spoof_gtp} & \ding{54} & & \ding{54} \\ \hline
PFCP in GTP \cite{pfcp_in_gtp} & \ding{54} & & \\ \hline

\end{tabular}
\caption{Mapping of labeled attacks to their detection dimensions}
\label{tab:attack_dim}
\end{table}

% Attaque SOTA
Our objective is to demonstrate that our approach can detect complex attacks spanning multiple detection dimensions. We therefore implemented seven attacks from the literature, covering complementary detection dimensions as well as different attack surfaces and protocols. These include attacks targeting PFCP for session management, such as establishment, deletion, and modification floods\cite{pfcp}. We also consider spoofing and redirection attacks targeting the GTP-U protocol \cite{spoof_gtp,pfcp_in_gtp}, related to how traffic is encapsulated toward the UPF and subsequently handled by it. Finally, we include scanning attacks targeting the 5G network architecture through HTTP/2 API calls\cite{5gad}.

% Mes attaques
We additionally introduced three original attacks to further diversify the considered attack scenarios. First, we implemented a Man-in-the-Middle (MITM) attack, covering both its setup phase, which generates network discovery, NF additions, and NF removals, and its subsequent operation, which redirects traffic. We also introduced API fuzzing, which requires the model to learn from a diverse range of benign HTTP/2 traffic in order to identify anomalous messages. Finally, we implemented SEID fuzzing, which represents a more challenging variant of session modification attacks, as the modified messages remain semantically legitimate. For each attack, we identified the detection dimension or dimensions that need to be considered for its detection. These mappings are summarized in Table~\ref{tab:attack_dim}. Although our attacks do not cover every existing 5G protocols and entities, we argue that they provide sufficient diversity in terms of attack surfaces and detection dimensions to serve as a proof of concept for the anomaly detection.  

\begin{table}[t]
\centering
\setlength{\tabcolsep}{3pt}
\begin{tabular}{|l|c|c|c|c|c|c|c|c|c|}
\hline
\diagbox[width=4.5cm]{\textbf{Attack procedure}}{\textbf{Benign procedure}}
& \rotatebox{90}{\textbf{Register UE}}
& \rotatebox{90}{\textbf{Set UE Idle}}
& \rotatebox{90}{\textbf{Uplink Wake}}
& \rotatebox{90}{\textbf{Downlink Wake}}
& \rotatebox{90}{\textbf{Restart Session}}
& \rotatebox{90}{\textbf{Deregister UE}}
& \rotatebox{90}{\textbf{Add NF}}
& \rotatebox{90}{\textbf{Remove NF}}
& \rotatebox{90}{\textbf{User Traffic}}
\\ \hline
Applicative MiTM & & & & & & & \ding{54} & \ding{54} & \\ \hline
Applicative Scan & & & & & & & \ding{54} & \ding{54} & \\ \hline
API Fuzzing & \ding{54} & \ding{54} & \ding{54} & \ding{54} & \ding{54} & \ding{54} & \ding{54} &  \ding{54} & \\ \hline
Session Establishment Flood & \ding{54} & \ding{54} & & & \ding{54} & & & & \\ \hline
Session Deletion Flood & & & & & & \ding{54} & & & \\ \hline
SEID Fuzzing & \ding{54} & \ding{54} & \ding{54} & \ding{54} & \ding{54} & & & & \\ \hline
Session Modify Drop & \ding{54} & \ding{54} & \ding{54} & \ding{54} & \ding{54} & & & & \\ \hline
Session Modify Duplicate & \ding{54} & \ding{54} & \ding{54} & \ding{54} & \ding{54} & & & & \\ \hline
Uplink Spoofing & & & & & & & & & \ding{54} \\ \hline
PFCP in GTP & & & & & & & & & \ding{54} \\ \hline
\end{tabular}
\caption{Mapping of labeled attacks to the benign procedures that provide comparable traffic for their detection.}
\label{tab:attack_vs_benign}
\end{table}

% Bénin doit être comparable
For an attack to be reliably detected, it is essential to have benign traffic of comparable nature, so that the two can be discriminated based on the specific characteristics that make the attack anomalous. For instance, comparing PFCP attacks exclusively against benign HTTP/2 traffic would not provide a meaningful basis for detection. However, this aspect is not systematically considered in existing works. In our dataset, we explicitly associate each attack with benign procedures that generate comparable traffic, as summarized in Table~\ref{tab:attack_vs_benign}.

% Contenu bénin
Our benign traffic includes procedures commonly used in the state of the art, such as UE registration and deregistration, which generate a wide variety of session-management traffic relevant to PFCP attacks. Combined with procedures that modify UE states, they also generate a diverse range of HTTP/2 messages required to detect our fuzzing attacks. We also implemented benign procedures not considered in previous works. These include NF addition and removal, which generate legitimate traffic comparable to that produced during MITM and scanning attacks. Similarly, we also included user-plane message transmission, which generates legitimate GTP-U encapsulation and therefore provides benign traffic against which GTP-U attacks can be identified. Overall, our dataset contains nine types of benign procedures, whose correspondence with the considered attacks is summarized in Table~\ref{tab:attack_vs_benign}.

\begin{figure}[ht]
\centering
\includegraphics[width=\linewidth]{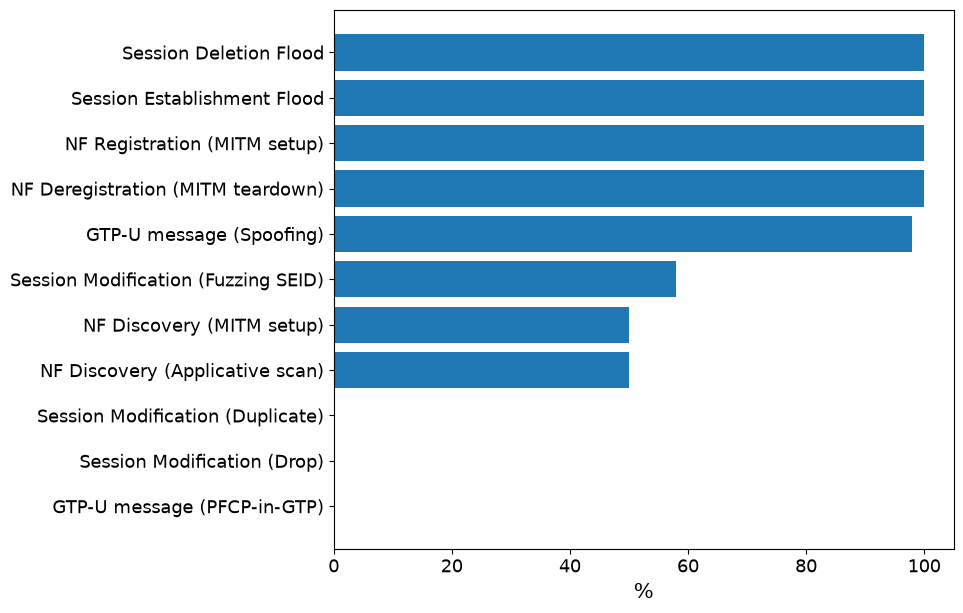}
\caption{Ratio of benign messages semantically identical to attack messages of the same nature}
\label{fig:jacard}
\end{figure}

% Analyse qualité : Sem
We subsequently analyzed the dataset to verify that attacks were detectable only along their characterized dimensions. For semantic anomalies, we computed the Jaccard distance between the protocol fields of attack messages and comparable benign messages. A distance of zero indicates semantically identical messages. Figure~\ref{fig:jacard} reports, for each attack, the proportion of benign messages of the same nature with zero distance to an attack message. Only PFCP in GTP and Session Modification have no identical benign counterpart. API fuzzing was analyzed separately by API call type, with 93\% of its messages having no identical benign counterpart.

\begin{figure}[ht]
\centering
\includegraphics[width=\linewidth]{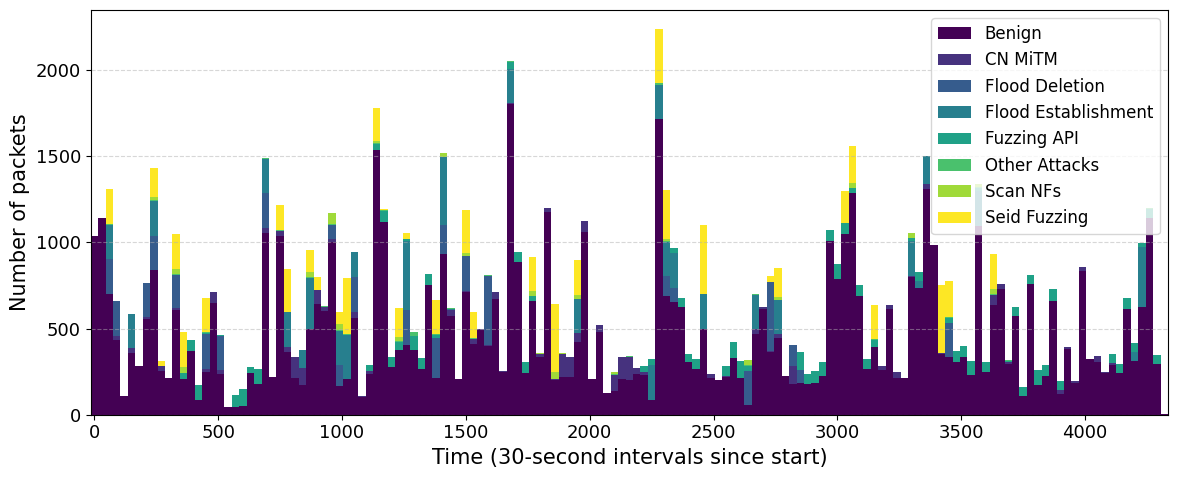}
\caption{Number of packets of different types per 30-second interval}
\label{fig:time}
\end{figure}

% Analyse qualité : Tempo
For temporal anomalies, we analyzed message distributions over time. As described above, benign and attack traffic are interleaved and attack times are randomized, preventing simple temporal isolation. Figure~\ref{fig:time} shows that repetitive attacks generally produce identifiable bursts. Scanning is an exception, as it requires relatively few packets, while API fuzzing is more dispersed because its execution takes longer. MITM also generates high message volumes due to relayed benign traffic. Overall, temporal patterns provide useful detection signals but are insufficient for some attacks.

% Analyse qualité : Topo
Finally, for topological anomalies, we searched the dataset for the specific patterns characterizing these attacks and verified that they occur exclusively in anomalous traffic. For MITM, we searched for a redirection pattern in which an NF forwards an API request to another NF and found approximately 4,500 occurrences due to the relayed benign traffic. For user spoofing, we searched for GTP-U requests with inconsistent external IP addresses between the request and response. This pattern occurred 72 times, compared with 92 attack executions, with the difference attributed to packet losses.

% Conclusion & Liens externes
Overall, this analysis demonstrates that the attacks in our dataset can only be detected according to the dimensions we defined for each of them, thereby demonstrating that our dataset is well suited for evaluating our formalism. Interested readers can refer to the detailed description of the attacks and benign procedures, as well as the data generation process and dataset analysis, available here\cite{CTD5G}. The implementation of our dataset generation algorithm is also publicly available on our GitHub repository\footnote{https://github.com/thomas-hoger/CTD-5G}, together with detailed documentation on how to reproduce the dataset generation process and adapt it to include additional attacks. Finally, the packet capture used in this paper is available here\footnote{doi.org/10.5281/zenodo.15853959}.

\subsection{Classification Metrics}

To evaluate the detection performance of our models, we use standard classification metrics. Our models output a probability score in $[0,1]$, where values close to 0 indicate benign behavior and values close to 1 indicate anomalous behavior.

For a given classification threshold, the TPR (True Positive Rate) measures the proportion of correctly detected attacks, while the FPR (False Positive Rate) measures the proportion of benign instances incorrectly classified as anomalous:
\[
\text{TPR} = \frac{TP}{TP+FN} \qquad
\text{FPR} = \frac{FP}{FP+TN}
\]

We aim to maximize TPR while minimizing FPR. To evaluate model discrimination independently of a specific threshold, we use the ROC curve, which plots TPR against FPR for all possible thresholds. We summarize the model's performance using the AUC (Area Under the Curve), where an AUC of $0.5$ corresponds to random classification, while an AUC of $1$ indicates perfect discrimination.

\subsection{Results}

\begin{figure}[!ht]
\centering
\includegraphics[width=0.9\linewidth]{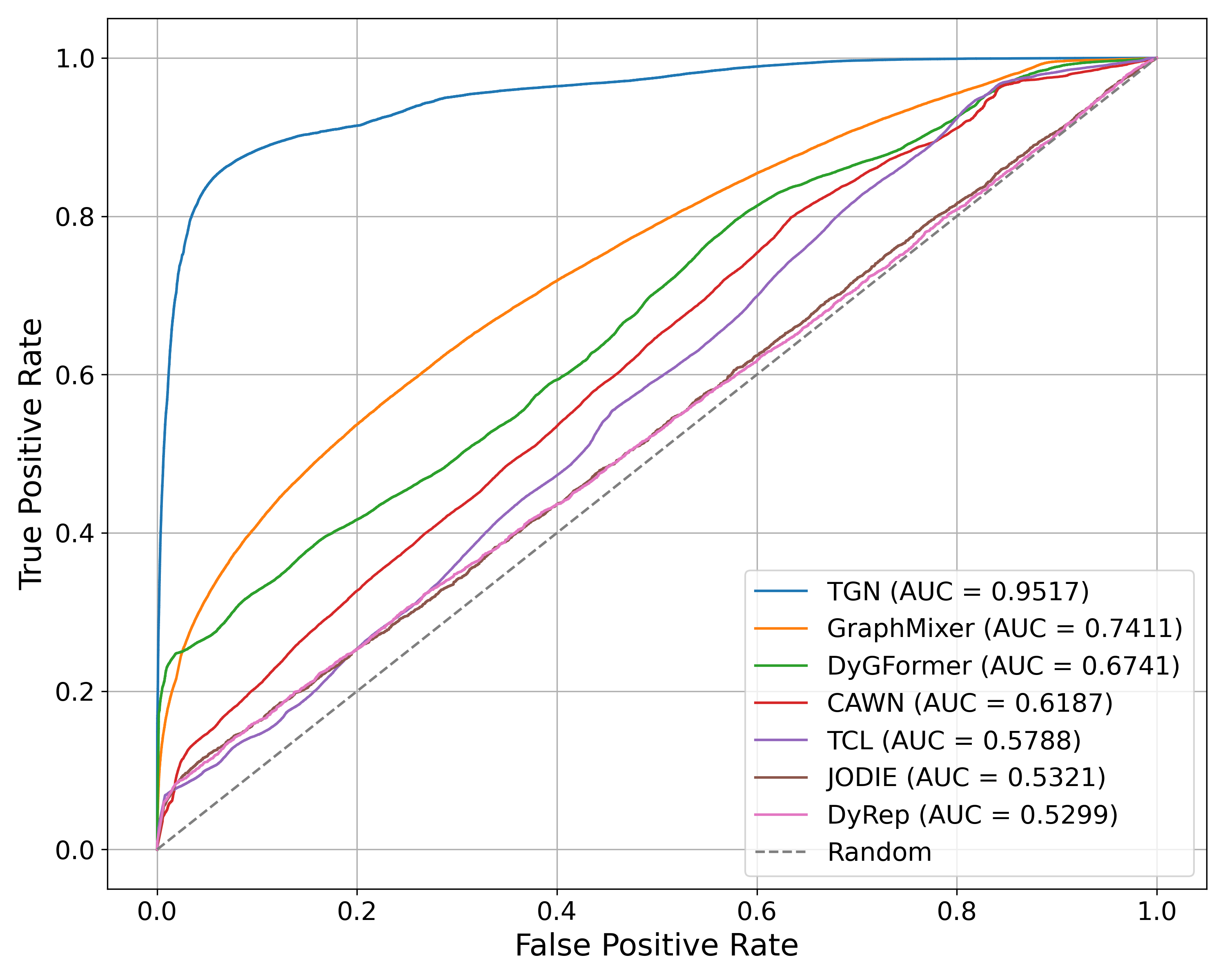}
\caption{ROC curves for the different supervised link prediction models.}
\label{fig:roc_sl}
\end{figure}

After training the seven state-of-the-art models presented in Section~\ref{methodology}, we obtained the ROC curves for each evaluated model, which are presented in Figure~\ref{fig:roc_sl}. Among the different models, TGN achieves the best performance, with an AUC of 0.95, substantially outperforming the other. This result is particularly encouraging and demonstrates the potential of the proposed formalism for supervised anomaly detection. While these results are promising, they should be interpreted cautiously in the context of attack detection. Indeed, even a low false-positive rate can generate a significant number of alerts when processing large volumes of traffic, while false negatives correspond to undetected attacks. Although the current performance is not yet sufficient for direct operational deployment, these results demonstrate the relevance of the proposed formalism for supervised anomaly detection.

\begin{table}[t]
\centering
\small
\begin{tabular}{lcc}
    \hline
    \textbf{Model} & \textbf{AUC} & \textbf{Packet/s} \\
    \hline
    JODIE      & 0.5321 & 90.63 \\
    DyRep      & 0.5299 & 41.20 \\
    CAWN       & 0.6187 & 138.43 \\
    DyGFormer  & 0.6741 & \underline{855.73} \\
    TCL        & 0.5788 & 151.64 \\
    GraphMixer & 0.7411 & 76.23 \\
    TGN        & \underline{0.9517} & 22.79 \\
    \hline
\end{tabular}
\caption{Packets per second (Packets/s) and AUC for the different evaluated models.}
\label{tab:supervise}
\end{table}

However, good detection performance alone is not sufficient when the objective is to perform real-time detection. Table~\ref{tab:supervise} therefore compares the performance and execution time of the evaluated models to assess whether their throughput is sufficient for real-time processing. Despite achieving the best detection performance, TGN is also the slowest model, processing approximately 23 packets/s. During dataset generation, we observed an average traffic rate of 29 packets/s, meaning that TGN is not yet capable of real-time processing. However, this gap is relatively small, and the main cause of the limited throughput appears to be the DyGLib implementation, which is not optimized for execution efficiency and performs inefficient memory allocations. The execution time could therefore be significantly reduced either by optimizing the implementation or by simplifying the model.

\section{Interpretation}
\label{interpretation}

As discussed, beyond supporting the three detection dimensions, our main contribution lies in the interpretability of the proposed formalism. In our approach, the model associates each graph edge with a probability reflecting its likelihood of being legitimate. These probabilities can then be directly projected onto the graph structure, highlighting the elements that contributed to the model decision.

\begin{figure}[!ht]
\centering
\includegraphics[width=1\linewidth]{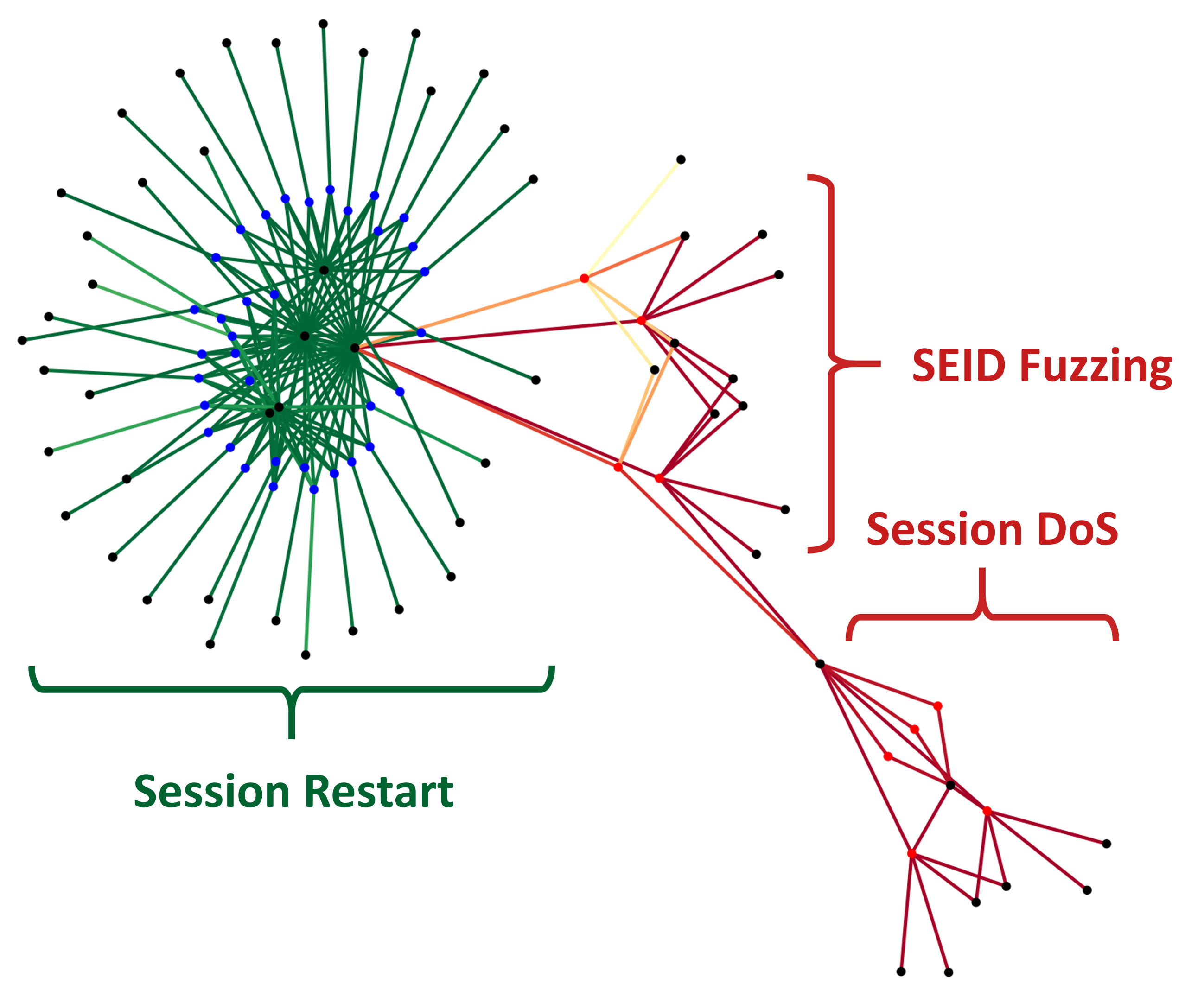}
\caption{Static graph representation of anomaly predictions over observed packets. Red edges indicate a high anomaly probability, while green edges indicate a low anomaly probability. Benign \textit{central nodes} are represented in blue and attacks by red nodes. The black nodes represent \textit{feature nodes}, which are not inherently benign or anomalous.}
\label{fig:interpretable}
\end{figure}

Figure~\ref{fig:interpretable} illustrates a static graph representation of a real scenario in which benign messages associated with a session restart procedure coexist with anomalies resulting from Session DoS and SEID Fuzzing attacks. Notably, attack-related edges are not all immediately classified as highly anomalous. For SEID Fuzzing, the first attack packets appear in orange, reflecting initially low suspicion based on their semantics. Subsequent packets introduce a temporal anomaly, increasing the model confidence in the attack. In contrast, Session DoS packets are immediately identified as highly anomalous due to their abnormal semantic characteristics.

\section{Conclusion and Future Work}
\label{conclusion}

Most existing approaches to 5G anomaly detection focus primarily on the semantic content of individual messages, while overlooking the sequential and topological context in which interactions occur. This limits their ability to capture complex and multi-stage attacks. In this work, we proposed a graph-based formalism that jointly represents the semantic, sequential, and interaction dimensions of 5G control-plane traffic. By explicitly modeling both the internal structure of packets and the temporal interactions between network entities, our approach provides a more expressive representation of 5G exchanges.

This work demonstrated the effectiveness of the proposed formalism in a supervised learning setting. Future work will investigate unsupervised approaches, which are more challenging to balance but may better address previously unseen threats. Regardless of the learning paradigm, the main advantage of our approach lies in its unified and highly interpretable representation, which can support the analysis of both known and novel anomalies.

\section*{Acknowledgments}
We would like to thank the PEPR program ``Networks of the Future`` of the Agence Nationale de la Recherche (ANR) for supporting our work as part of the France 2030 investment plan. In particular, we would like to thank the NF-HiSec project, funded under reference ANR-22-PEFT-0009, which played a key role in the realization of this work.

\bibliography{references}

\vspace{2pt}

% Print the glossary with one of the mcol styles.
%\printglossary[type=\acronymtype]

%\appendix

%\input{appendix.tex}

%\vspace{12pt}

\end{document}